\documentclass[aps,prl,superscriptaddress,showpacs,twocolumn,floatfix]{revtex4}
\usepackage{graphicx}
\usepackage[usenames,dvipsnames]{color}

\newcommand{\PRL}{{ Phys. Rev. Lett. }}

\newcommand{\etal}{{\em et al.}}
\newcommand\JLab{Thomas Jefferson National Accelerator Facility, Newport News, VA 23606 }
\begin{document}
\preprint{APS/123-QED}
\title{On a sidereal time variation of the Lorentz force}
\author{B.~Wojtsekhowski}
\affiliation{\JLab}
\date{\today}

\begin{abstract}
We consider a search for a sidereal time variation of the beam trajectory 
in the cyclotron motion in a static magnetic field. 
The combination of two beams moving in opposite directions could allow a test 
of the radius stability with sensitivity approaching $10^{-10}$.
Such a level of variation of the Lorentz force, if it exists, would require 
a speed of light anisotropy on the level of $10^{-18}$.
\end{abstract}


\maketitle

The speed of light has been the subject of experimental and theoretical studies for hundreds of years.
The constancy of the speed of light has been the keystone of modern physics since the formulation
of the theory of special relativity (SR) by A.~Einstein.
The postulate of SR is supported by the precision measurement of the isotropy of the speed
of light by Michelson\&Morley~\cite{MM1887}, who, in fact, measured a difference in the speed of light in orthogonal
directions averaged in an apparatus over two opposite directions (so-called two-way speed) with 
a relative precision of $10^{-5}$. 
The current best limit for this type of anisotropy ($\Delta c_{_2}/c \leq 10^{-18}$) was obtained 
in the resonating cavities experiment~\cite{Tobar2014}.
The anisotropy of one-way speed characterizes the difference in the speed of light 
in two different directions (e.g. in opposite directions) and is currently constrained 
on the level of $\Delta c_{_1}/c \leq 10^{-14}$~\cite{CAVITY2013, GRALL2010}.
In this paper we discuss the motion of a charged particle in a static transverse magnetic 
field as a sensitive test of a sidereal time variation of the Lorentz force (VLF) as well as 
possible consequences for the speed of light anisotropy (SLA).
We are detailing here a concept first discussed in Ref.~\cite{BW2014}.

The proposed method uses a static magnetic field in two ways: the first, for measurement of the particle momentum
via the radius of its trajectory, and the second, for changing the direction of the particle motion.
The changing direction allows a prompt search for the VLF in an additional momentum 
measurement before the parameters of the magnetic system and particle energy drift
due to the apparatus instabilities.
Finally, the phase of the sidereal time is used as usual in the signal processing for suppression
of the noise and systematics with different frequencies.

The experiment proposed here will test the isotropy of the maximum speed of electrons/positrons, 
but we refer to the speed of light for convenience.
The experiment using an electron beam in combination with the electron/positron version presented here
would allow a test of the CPT symmetry in an electron/positron sector for the maximum attainable speed.

\paragraph{The speed of light anisotropy and the Lorentz force:}
A number of predictions of the theory of special relativity were tested with very high accuracy, e.g.
the Doppler effect~\cite{GSI}.
It is also predicted that a trajectory radius of the charged particle in a cyclotron motion is 
independent of the trajectory orientation in space. 
To our knowledge, this prediction has not been tested accurately. 
The cyclotron motion allows us to evaluate the interconnection between the SLA and the Lorentz force.
Indeed, a vector of the conventional Lorentz force~\cite{LANDAU} is proportional to the vector product 
of the charged particle speed $\vec v$ and magnetic field $\vec B$: $\vec v \times \vec B$.
In the framework of SR the linear electrodynamics
leads to a constancy of the absolute values of both the speed $|\vec v|$ 
and momentum $|\vec p|$ during the charged particle motion in a static magnetic field.
However, in the case of the SLA, the value of the particle momentum for a given value of speed $|\vec v|$ is
dependent on the angle between the direction of charged particle motion $\vec v/|\vec v|$
and the preferable direction $\vec e$ of the SLA.
The momentum or speed (or both of them) should vary during the cyclotron motion if the SLA is non-zero.
The speed of light anisotropy and conventional Lorentz force are not compatible at
the level needed for the analysis of the cyclotron motion.

\paragraph{Directional variation of the momentum:}
The possible generalized dispersion equation $E^2 = m^2 + p^2 + f \cdot \vec e \cdot \vec p + ...$
(where $f$ is a model parameter) used in different theoretical constructions of 
the quantum gravity (see e.g. the review~\cite{MATT2005}) has the term 
$\vec e \cdot \vec p$ which requires a directional variation of the absolute value of momentum
if energy conservation is assumed.
Instead of a 180$^\circ$ turn by a static magnetic field we could consider a turn around in
an elastic scattering of the probe particle from a hypothetical massive object.
The particle momenta before and after scattering could be measured via deflection in a magnetic field.

\paragraph{Modification of the Lorentz force:}
There are only three vectors in the cyclotron motion problem in the presence of the SLA:
$\vec v$, $\vec B$, and $\vec e$, which could be used for construction of the force vector.
Due to the axial nature of the magnetic field, the $\vec B$ should be used in a combination $\vec v \times \vec B$.
The resulting form of the force could be $\vec v \times \vec B + \alpha \cdot \vec e \, (\vec e \cdot [\vec v \times \vec B])$,
where the $\alpha$ is a new parameter related to the SLA value.
The time-reversal symmetry is violated by such an additional term.
Over a period of revolution, the value of the particle momentum is restored.

\paragraph{Sensitivity of the cyclotron motion to the SLA:}
Assuming that the absolute value of speed $|\vec v|$ is constant in the cyclotron motion,
as it is in the case of the conventional Lorentz force, we can find the sensitivity
of the experiment to the variation of the speed of light.
We are considering a local radius $R$ of a charged particle trajectory in a static magnetic field. 
There is a large amplification of the $R$ variation relative to the speed of light variation:
$\Delta R/R = \gamma^2 \Delta c /c$, where $\gamma = 1/\sqrt{1-(v/c)^2}$ is a particle Lorentz factor. 
Currently we can use a beam of particles with the Lorentz factor on the level of $10^4$,
so sensitivity to $\Delta c/c$ is enhanced by eight orders of magnitude.

\paragraph{Experimental approach to the probing of the Lorentz force:}
A number of high energy particle accelerators are suitable for trajectory
measurements because of the available high intensity of particle beams and 
the high precision of beam position monitors (BPM)~\cite{ACC-BPM}.

The average momentum of the beam in the accelerator $p$ is defined by the magnetic field.
Within the accelerator acceptance, the beam coordinate $x$ in the dispersive plane 
relative to the nominal trajectory could be presented as $x(s) = x_\beta(s) + \eta(s) \cdot (\Delta p/p)$, 
where $s$ is the beam path along the trajectory, $\eta(s)$ is a dispersion function, and
$\Delta p$ is the deviation of the momentum from a nominal value.
In a typical electron/positron storage ring the damping time of the beam betatron oscillation
is well below one second, so we can disregard the $x_\beta$ term. 
The dispersion function is defined by a full set of magnetic elements in the beam trajectory
and could usually be found to a few percent accuracy having a typical value up to a few meters.
The $\eta$ function varies strongly along the beam path, allowing for measurement 
of the momentum deviation $\Delta p$ using the BPM data.

Using, for example the BPM developed for the International Linear Collider~\cite{ILC-BPM}, 
the beam position could be determined to 15~nm in one pass, which corresponds to 
the $\Delta p/p \sim 10^{-8}$.
BPMs with an accuracy of 1~$\mu$m are common (see e.g. Ref.~\cite{KEK-BPM})
and allow a $10^{-6}$ relative accuracy for the momentum measurement every time
the beam passes a few BPMs in an area of high dispersion.
Measurement over 100 seconds would provide a statistical precision of $10^{-10}$.

\paragraph{Ratio of the momenta in opposite directions:} 
A high statistical accuracy of the beam position measurement means that the limit on
experiment sensitivity will come predominantly from instabilities of the beam energy
and magnetic system.
The impact of the beam energy fluctuations on the precision of the VLF search
could be reduced by doing the momentum measurements at two opposite (or several different) 
directions of the beam motion during the same beam turn in the ring and 
constructing the ratio(s)~\cite{BW2014}.

\begin{figure}[!thb]
\centering
\includegraphics[width=0.4\textwidth]{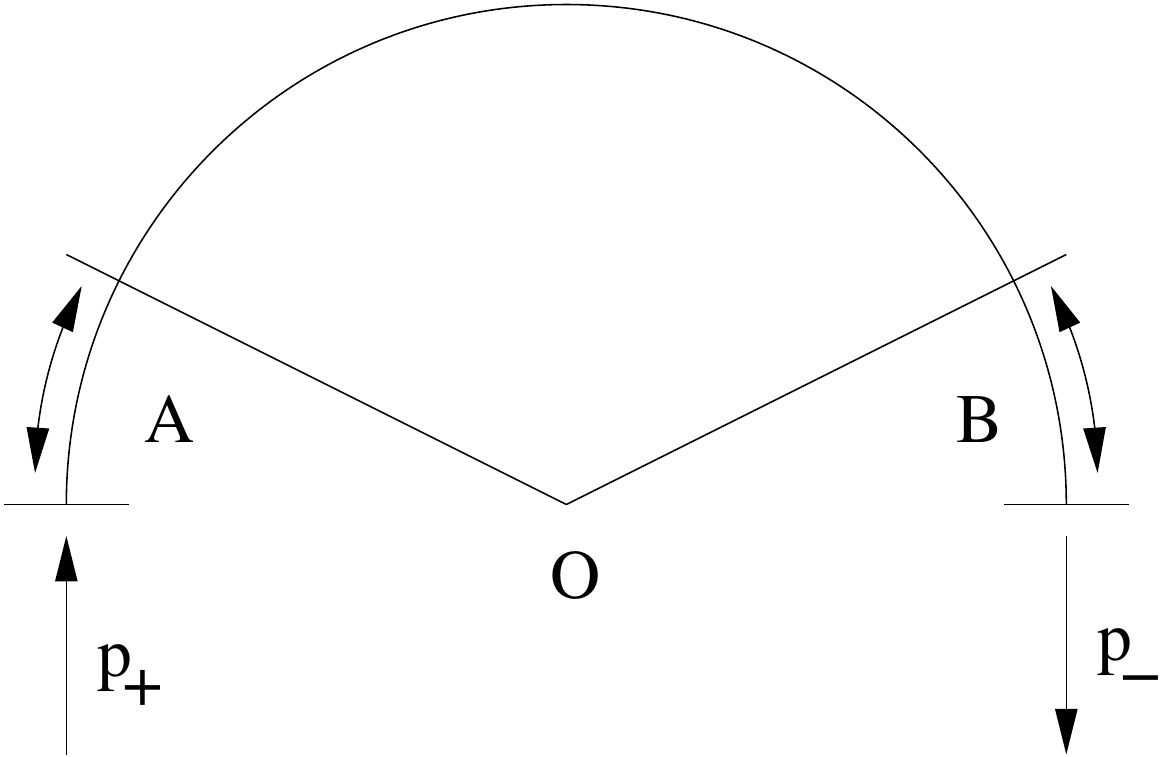} 
\caption{Diagram for measurement of the beam momenta with a 180$^\circ$ arc magnet.}  
\label{fig:arc}
\end{figure}

An illustrative example of this idea is the following:
Two beam momentum monitors 
(each of them includes a few BPMs in a region with a large $\eta$ function)
at opposite sides of a 180$^\circ$ bending magnet allow one to measure the ratio 
of the particle momenta $p_+, \, p_-$ moving in opposite directions, 
$R = p_+ / p_-$ (see Fig.~\ref{fig:arc}).

Assuming that there are no acceleration elements between those two momentum 
measurements and a small calculable correction on the radiative energy loss, 
such a ratio should be stable even when the actual beam energy varies.
The value of $R$ will be close to 1 with some deviation mainly due to the systematic 
uncertainty of the momentum monitors calibration.
The configuration could have a large portion of the arc used as a momentum monitor
(the area A (B) shown in Fig.~\ref{fig:arc}) because it would allow 
better sensitivity in spite of the angle's being smaller than 180$^\circ$ 
between the two momentum monitors.
The systematics of the $R$ measurement are strongly suppressed when the measurements in 
both monitors perform synchronously.

\paragraph{Two beams in one ring:}
The signal of interest varies slowly with sidereal time.
It is not an easy task to achieve a level of stability as good as $10^{-10}$ over a
24-hour period due to the storage ring temperature variations and drifts of the measuring instruments.
However, in several storage rings there is a possibility of making measurements with an electron
beam and a positron beam turning in opposite directions.
Two beams could be present in the ring simultaneously or within a short time one 
after another without changes of the magnetic field in the ring.
A two-beam measurement would allow a compensation of the drift of the  momentum monitor 
characteristics via normalization of the electron ratio $R^e = p_+^e / p_-^e$ to the positron 
ratio $R^p = p_+^p / p_-^p$.
The double ratio $R_{e/p} = R^e/R^p$ would be immune to most instabilities.
This double ratio also has twice the sensitivity to the directional variation of the speed of light.
Such a measurement could be performed at the storage rings CESR and VEPP-4.
They operate at an energy of 5 GeV and have a small energy loss for
synchrotron radiation of $10^{-4}$ per turn.
The stability of the storage ring magnetic system could be as good as $10^{-7}$ over 
a few weeks~\cite{g-2}.
Operation with one bunch per beam should allow recording of the trajectories of both beams
synchronously, so the effective recalibration time of the instrument is of a few milliseconds 
from which the required stability of $10^{-10}$ of magnets and geometry is likely to be achievable.

\paragraph{Electron beam experiment:}
Doing the measurement with one beam is also important because it allows a test 
of the CPT theorem for the maximum attainable speed of an electron and a positron.
The JLab CEBAF accelerator~\cite{CEBAF} has the most advantageous parameters 
for the electron beam case because its beam has a high current of 100 $\mu$A,
a relative energy spread of a few 10$^{-5}$, and 
geometrical emittance of 10$^{-9}$ m$\times$rad (in spite of significant broadening
due to radiation losses in the last few turns).
The accelerator sections are separated by 180$^\circ$ arcs, 
and the beam recently reached $\gamma \sim 2 \times 10^4$,
the highest Lorentz factor among the currently operating accelerators. 
The absolute value of the beam energy in a linear accelerator 
could vary, but this does not impact the proposed investigation because 
only the ratio of the beam momenta at opposite sides of the arc needs to be stable.
In addition, the value of the anisotropy could be constrained
in 10 different arcs with increasing beam energy.
A small accelerating cavity in the middle of a 180$^\circ$ arc could be used 
for a direct test of the measurement sensitivity.

There are a number of storage rings which could
be used for the proposed measurement.
In the case, where only one beam type is available (e.g. an electron beam), 
the effect of the variation of the average beam energy on the trajectory 
can not be compensated for. 
However, the beam energy could be measured by means of the resonance 
spin depolarization method~\cite{POL} and its' slow variation could be taken into account.

An interesting option of the electron beam experiment could be realized
with the all-electrical storage ring proposed for the proton EDM search
experiment, where deflection of the beams is arranged via
a transverse static electric field~\cite{YANNIS}.
In this case we can use two electron beams circulating in 
opposite directions with a Lorentz factor of $10^3$.

\paragraph{Implications of the Lorentz force experiment:}
It would be interesting to find out the implication of the experimental
limit on the VLF for the limit of the speed of light anisotropy. 
However, at present such a type of connection has not been formulated.
It is not excluded that the charged particle trajectory in a static field is not
sensitive to the SLA in spite of modification of the Lorentz force. 
At same time we can formulate reversed implications:
The existing limit on the SLA of $10^{-14}$ restricts the value of 
possible VLF to below $10^{-6}$ for the beam energy of 5 GeV,
and if the VLF signal exists it will naturally require a non-zero SLA.
The current limit on the VLF is 10,000 times less stringent than the
sensitivity of the proposed measurement.

\acknowledgments

The author is grateful to D.~Rubin and V.~Zelevinsky for helpful discussions.
This work was supported by the U.S. Department of Energy. 
Jefferson Science Associates, LLC, operates Jefferson Lab for 
the U.S. DOE under U.S. DOE contract DE-AC05-060R23177.

\end{document}